\documentclass[aps,pra,twocolumn,showpacs]{revtex4}

\usepackage{graphicx}
\usepackage{amsmath,amssymb,amsthm,bbm,latexsym}
\usepackage{amsfonts}

\begin{document}

\title{The stabilizer dimension of graph states}

\author{D.H. Zhang}
\affiliation{Beijing National Laboratory for Condensed Matter
Physics, and Institute of Physics, Chinese Academy of Sciences,
Beijing 100190, China}

\author{H. Fan}
\affiliation{Beijing National Laboratory for Condensed Matter
Physics, and Institute of Physics, Chinese Academy of Sciences,
Beijing 100190, China}

\author{D.L. Zhou}
\affiliation{Beijing National Laboratory for Condensed Matter
Physics, and Institute of Physics, Chinese Academy of Sciences,
Beijing 100190, China}

\begin{abstract}
The entanglement properties of a multiparty pure state are invariant under local unitary transformations. The stabilizer dimension of a multiparty pure state characterizes how many types of such local unitary transformations existing for the state. We find that the stabilizer dimension of an $n$-qubit ($n\ge 2$) graph state is associated with three specific configurations in its graph. We further show that the stabilizer dimension of an $n$-qubit ($n\ge 3$) graph state is equal to the degree of irreducible two-qubit correlations in the state.  
\end{abstract}


\maketitle

\section{Introduction}

Entanglement is a useful resource in quantum information and quantum computation. How to classify and quantify entanglement in a multiparty quantum state is a fundamental theoretical problem. Although extensive investigations have been made in this direction \cite{Hor08}, we still lack a general characterization of multiparty entanglement \cite{Ple07}. 


A basic observation on multiparty entanglement is that any two multiparty quantum states that can be transformed into each other have the same entanglement properties \cite{Lin98,Lin99}. An orbit in the Hilbert space is defined as a set of states, in which any two states can be transformed into each other by a local unitary transformation. In this sense, one orbit represents one type of entanglement, and different orbits construct a classification of entanglement. One important property of an orbit is its dimension, \textit{i.e.}, the number of parameters needed to describe the position on the orbit. Obviously, these parameters describe local properties, and they are irrelevant of the degree of entanglement.

To obtain the dimension of an orbit, we can study its stabilizer group for any point on the orbit \cite{Car00}. Here a point on an orbit refers to a state in the Hilbert space. The stabilizer group for any point on an orbit is a subgroup of the group of local unitary transformations, whose element leaves the point invariant under its action. The dimension of the stabilizer group is the number of independent parameters in such loal unitary transformations, which is important in classifying different types of orbits.  In fact, the sum of the dimension of an orbit and the dimension of its stabilizer group is the dimension of the group of local unitary transformations, thus the dimension of an orbit can be obtained by investigating the dimension of its stabilizer group.

Much progress on the properties of orbits and their stabilizer groups has been achieved. In Ref. \cite{Aci01}, Ac\'{i}n \textit{et al.} give a generalized Schmidt decomposition of a three-qubit state in terms of five nonlocal parameters, which leads to a complete classification of three-qubit pure states. A more general discussion on the generalized Schmidt decomposition of a multiparty pure state is given in Ref. \cite{Car02}. In Ref. \cite{Kus01}, Ku\'{s} and \.{Z}yczkowski discuss similar idea for two-party mixed states. Very recently, Lyons \textit{et al.} investigate the related problem on the maximum stabilizer dimension of a pure multiqubit state in a series of papers \cite{Lyo05,Lyo07,Lyo08-1,Lyo08-2}. 

A multiqubit graph state \cite{Bri01} is a typical pure state with particular entanglement properties, and can be used as a universal entanglement resource in one way quantum computer \cite {Rau01,Rau03}. Further more, it is conveniently described in the stabilizer formalism, which is proposed in quantum error correcting codes \cite{Got97}.  Here we emphasize that, in the above contexts, we have used two stabilizer formalisms, where the former is a subgroup of local unitary transformation, and the latter is a subgroup of Pauli group. 

The entanglement properties of graph states have been extensively explored in literature \cite{Hei06,Zho08}. As far as we know, however, we still lack a good understanding on the basic question: what is the stabilizer dimension of any multiqubit graph state?  In this article, we will give a satisfactory answer to this question, and more precisely, we will obtain the analytical result for the stabilizer dimension of an arbitrary multiqubit graph state. 

This article is organized as follows. In Sec. II, we will introduce the basic concepts on graph states. In Sec. III, we will show how to obtain the stabilizer dimension for arbitrary graph states. In Sec. IV, we will prove that the stabilizer dimension of an $n$-qubit ($(n\ge 3$) graph state is equal to the degree of irreducible two-qubit correlation obtained in Ref. \cite{Zho08}. Finally, we will discuss one possible extension of our results and give a summary.

\section{Notation on graph states}

An $n$-qubit graph state can be regarded as a generalization of a $2$-qubit Bell state and a $3$-qubit GHZ state. Literally, a graph state is a state associated with a graph $\mathfrak{G}$, where the graph $\mathfrak{G}$ is defined by a set of vertices $\mathcal{V}$ and a set of edges $\mathcal{E}$. Given any graph, we can define a quantum state by the following rules. Each vertex in the graph denotes one qubit, and each edge between two vertices in the graph represents a unitary transformation on the two qubits corresponding to the edge. More precisely, the graph state corresponding to a graph $\mathfrak{G}$ can be written as
\begin{equation}
|\Psi_{\mathfrak{G}}\rangle=\prod_{(i,j)\in \mathcal{E}} U^{(i,j)}\bigotimes_{k\in\mathcal{V}} |+\rangle_{k} \label{eqpsi}
\end{equation}
with the two-qubit unitary transformation
\begin{equation}
U^{(i,j)}=\frac {I^{(i)} I^{(j)} + Z^{(i)} I^{(j)} + I^{(i)} Z^{(j)} - Z^{(i)} Z^{(j)}} {2}, \label{equij}
\end{equation}
where $I^{(i)}$ is the identity operator for the $i$-th qubit, $X^{(i)}$, $Y^{(i)}$, and $Z^{(i)}$ are the three Pauli matrices of $i$-th qubit. The state vector $|+\rangle_{k}$ is the eigen state with eigenvalue $+1$ of the Pauli matrix $X^{(k)}$. For the sake of simplicity, we will omit the identity operators in the following expressions. Obviously, if a graph includes several disconnected parts, then the corresponding graph state is a product state of the states of these unconnected parts. Thus we can investigate the entanglement properties of the graph state by studying the states of these unconnected parts respectively. Therefore, without losing any generality, we will only consider the states associated with a connected graph throughout this article.

An $n$-qubit graph state defined by Eqs. (\ref{eqpsi}) and (\ref{equij}) can be represented in the stabilizer formalism. The generators of the stabilizer group for the graph state $|\Psi_{\mathfrak{G}}\rangle$ are given by
\begin{equation}
g_{i}=X^{(i)}\prod_{j\in \mathcal{N}(i)}Z^{(j)}, \label{eqgi}
\end{equation}
where the set of neighbours of vertex $i$ is defined as 
$\mathcal{N}(i)=\{j\in \mathcal{V}\;|\;(i,j)\in\mathcal{E}\}$.  The stabilizer group for an $n$-qubit graph state is denoted as $\mathcal{G}_{n}$, which is an Abelian subgroup of $n$-qubit Pauli group. The set of the generators of the stabilizer group is denoted as $\mathfrak{g}_{n}$, whose cardinality is $n$ but whose elements are not uniquely determined. The set $\{g_{i}\}$ is only one possible choice for $\mathfrak{g}_{n}$. 

Another useful choice of the set $\{g_{i}\}$ has been introduced in Ref. \cite{Zho08}. It can be defined naturally from the density matrix of the graph state $|\Psi_{\mathfrak{G}}\rangle$.  Using the given set of the generators (\ref{eqgi}), we can write the density matrix of $|\Psi_{\mathfrak{G}}\rangle$ in the following form
\begin{equation}
\rho_{\mathfrak{G}}=\frac {1} {2^{n}} \prod_{i=1}^{n}(1+g_{i})=\frac {1} {2^{n}} \sum_zS_{z},
\end{equation}
where
\begin{equation}
S_{z}=\prod_i(g_i)^{z_i}=\pm\prod_{i} O^{(i)} \label{eqsz}
\end{equation}
are the $2^{n}$ elements in the group $\mathcal{G}_{n}$ with $O^{(i)}\in\{ X^{(i)},Y^{(i)},Z^{(i)},I^{(i)}\}$, and $z=(z_1,z_2,\ldots,z_n)$ with $z_i\in\{0,1\}$ for $i\in\{1,2,\cdots,n\}$. Let $\mathrm{supp}(S_{z})=\{i|O^{(i)}\in\{X^{(i)},Y^{(i)},Z^{(i)}\}\}$. Obviously, $\mathrm{supp}(S_z)\subseteq \{1,2,\cdots,n\}$, and the less the cardinality of the support of an operator, the simpler the form of the operator. Therefore, the simplest form of the generators of the stabilizer group $\mathcal{G}_{n}$ is given by the rule: the sum of the cardinalities of the generators will be as small as possible. This can be done by introducing a series of subsets of $\mathcal{G}_{n}$ as $\mathcal{G}_{k}=\{S_{z}\in \mathcal{G}_{n} | \big|\mathrm{supp}(S_{z})\big|\le k\}$ for $k\in\{1,2,\cdots,n\}$. Then we can make the generators of the subsets satisfy $\mathfrak{g}_{k}\subseteq \mathfrak{g}_{k+1}$ for $k\in\{1,2,\cdots,n-1\}$. We remark that the subsets themselves are not neccessarily a group, but we can still define the generators of each subset such that its elements can be uniquely expressed as the product of the generators. Although the generators are not completely determined, the cardinality of the generators is.

\section{The stabilizer dimension of graph states}

In this section, we will consider the calculation of the stabilizer dimension for a graph state. The stabilizer dimension for an $n$-qubit pure state $|\psi\rangle$ can be calculated as follows. To define the stabilizer dimension of the state $|\psi\rangle$, we will find the Lie group that is composed by the local unitary transformations whose actions do not change the state $|\psi\rangle$. The dimension of the Lie group is defined as the stabilizer dimension of the state. It is easy to prove that the state $|\psi\rangle$ is the common eigenstates of all the elements of the Lie algebra for the Lie group with eigenvalues $0$. Because the dimension of the Lie algebra is the same as the dimension of its Lie group, it is convenient to calculate the dimension of the Lie algebra as the stabilizer dimension.    

In the $n$-qubit Hilbert space $\mathcal{H}=(C^2)^{\bigotimes n}$, the Lie algebra of the local unitary group can be written as 
\begin{equation}
g_{L}=\theta+\sum_{a=1}^{n} g_{L}^{(a)},
\end{equation}
where $\theta$ is a real parameter, and
\begin{equation}
g_{L}^{(a)}=t_{ax} X^{(a)}+t_{ay} Y^{(a)}+t_{az} Z^{(a)}.
\end{equation}
For the graph state $|\Psi_{\mathfrak{G}}\rangle$, we have
\begin{equation}
g_{L}| \Psi_{\mathfrak{G}}\rangle=0.\label{eqgl}
\end{equation}
Inserting Eq. (\ref{eqpsi}) into Eq. (\ref{eqgl}), we obtain
\begin{widetext}
\begin{equation}
\Big[\theta+\sum_{a\in\mathcal{V}}(t_{ax}\prod_{b\in \mathcal{N}(a)}Z^{(b)}-i t_{ay} Z^{(a)}\prod_{b\in \mathcal{N}(a)} Z^{(b)}+t_{az}Z^{(a)})\Big]\bigotimes_{k\in \mathcal{V}} | +\rangle_{k}=0.
\end{equation}
\end{widetext}

Because the operator $Z$ transforms the state $|+\rangle$ into $|-\rangle$, nonzero solutions of the above equation exist when at least two terms have the same $Z$ pattern. A direct conclusion is that $\theta=0$. We remark that $\theta$ is associated with $U(1)$ symmetry of the state, \textit{i.e.}, it is related with the global phase of the state.    

The solutions can be classified into three different cases. 

i) Only two terms from $X$ are nonzero, namely, we have
\begin{equation}
t_{ax}\prod_{c\in \mathcal{N}(a)}Z^{(c)}+t_{bx}\prod_{d\in \mathcal{N}(b)}Z^{(d)}=0.
\end{equation}
The solution of the above equation is $\mathcal{N}(a)=\mathcal{N}(b)$ and $t_{ax}=-t_{bx}$. The first equation implies that qubit $a$ and qubit $b$  have the same neighbor(s) as shown in FIG. 1. The second equation shows that the Lie algebra $g_{L}=X^{(a)}-X^{(b)}$. In addition, if there is such a configuration in the graph $\mathfrak{G}$, we will find that $X^{(a)}X^{(b)}$ is in the stabilizer group of the state $|\Psi_{\mathfrak{G}}\rangle$. 
\begin{figure}[htbp]
\begin{center}
\centering
\includegraphics[height=1.3in]{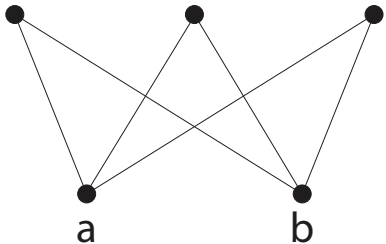}
\caption{(Configuration 1) Two vertices $a$ and $b$ have the same neighbor(s). Only the vertices and the edges related to $a$ and $b$ are shown.}
\end{center}
\end{figure}

ii) Only one term from $X$ and one term from $Z$ are nonzero, namely, we have
\begin{equation}
t_{ax}\prod_{c\in \mathcal{N}(a)}Z^{(c)}+t_{bz}Z^{(b)}=0.
\end{equation}
The solution of the above equation is $\mathcal{N}(a)=\{b\}$ and $t_{ax}=-t_{bz}$. The first equation implies that qubit $b$ is the unique neighbour of qubit $a$  as shown in FIG. 2. The second equation shows that the Lie algebra $g_{L}=X^{(a)}-Z^{(b)}$. In addition, if there is such a configuration in the graph $\mathfrak{G}$, we will find that $X^{(a)}Z^{(b)}$ is in the stabilizer group of the state $|\Psi_{\mathfrak{G}}\rangle$. 

\begin{figure}[htbp]
\begin{center}
\centering
\includegraphics[height=1.3in]{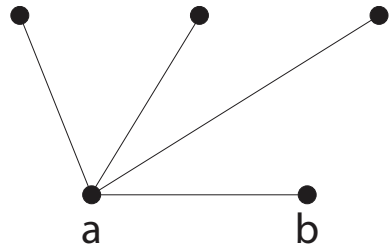}
\caption{(Confguration 2) One vertex $b$ has a unique neighbour $a$. Only the vertices and the edges related to $a$ and $b$ are shown.}
\end{center}
\end{figure}

iii) Only two terms from $Y$ are nonzero, namely, we have
\begin{equation}
t_{ay}\prod_{c\in \mathcal{N}(a)}Z^{(c)}Z^{(a)}+t_{by}\prod_{d\in \mathcal{N}(b)}Z^{(d)}Z^{(b)}=0.
\end{equation}
The solution of the above equation is $\mathcal{N}(a)\cup\{a\}=\mathcal{N}(b)\cup\{b\}$ and $t_{ay}=-t_{by}$. The first equation implies that qubit $a$ and qubit $b$ are neigbours, and all their other neighbours are the same. This configuration is shown in FIG. 3. The second equation shows that the Lie algebra $g_{L}=Y^{(a)}-Y^{(b)}$. In addition, if there is such a configuration in the graph $\mathfrak{G}$, we will find that $Y^{(a)}Y^{(b)}$ is in the stabilizer group of the state $|\Psi_{\mathfrak{G}}\rangle$. 

\begin{figure}[htbp]
\begin{center}
\centering
\includegraphics[height=1.3in]{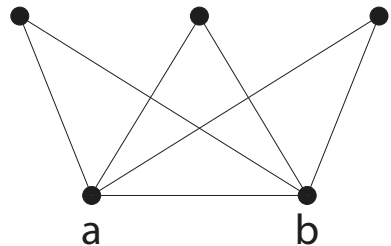}
\caption{(Configuration 3) Two vertices $a$ and $b$ are neigbours, and all their other neighbours are the same. Only the vertices and the edges related to $a$ and $b$ are shown.}
\end{center}
\end{figure}

Thus the stabilizer dimension of the graph state $|\Psi_{\mathfrak{G}}\rangle$ equals to the number of independent Lie algebras from the three configurations appearing in the graph $\mathfrak{G}$.

For example, a two-qubit graph state as shown in FIG. 4 (a), which is local unitary equivalent to the two-qubit Bell state, can be regarded as configuration $2$ and configuration $3$. Thus we find three independent generators of Lie algebra $g_{L1}=Z^{(1)}-X^{(2)}$, $g_{L2}=Z^{(2)}-X^{(1)}$, and $g_{L3}=Y^{(1)}-Y^{(2)}$. It implies that the stabilizer dimension of the two-qubit graph state is $3$.

Another example is an $n$-qubit tree-like graph state, which is local unitary equivalent to multi-qubit GHZ state. The graph has configuration $2$, thus we know that there are $n-1$ independent generators of Lie algebra, which are $\{g_{Li}=Z^{(1)}-X^{(i+1)},\;i\in(1,n-1)\}$. Therefore its stabilizer dimension is $n-1$.

\begin{figure}[htbp]
\begin{center}
\centering
\includegraphics[height=2.4in]{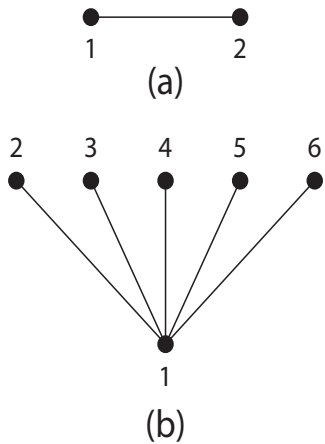}
\caption{(a) A two-qubit graph. (b) An $n$-qubit tree-like graph for $n=7$.}
\end{center}
\end{figure}

\section{The stabilizer dimension and irreducible two-qubit correlation}

In this section, we will show that the stabilizer dimension for an $n$-qubit ($n\ge 3$) graph state is equal to the degree of irreducible two-qubit correlation defined in Ref. \cite{Zho08}. 

First, we observe that if there is an element whose support's dimension is $2$ in the stabilizer group of the graph state $|\Psi_{\mathfrak{G}}\rangle$, then the two vertices in the support must be in one of the above three configurations. This can be proved as follows.

Assume that the element is denoted as $S_{z}$ whose support is $\{a,b\}$. Because it can be written in the form of Eq. (\ref{eqsz}),  we get $z_{i}=0$ for $i\notin \{a, b\}$ and 
\begin{equation}
\mathrm{supp}\Big({X^{(a)}}^{z_{a}}\prod_{c\in \mathcal{N}(a)}{Z^{(c)}}^{z_{a}}{X^{(b)}}^{z_{b}}\prod_{d\in \mathcal{N}(b)}{Z^{(c)}}^{z_{b}}\Big)=\{a,b\}. \label{eqsu}
\end{equation}
In the case of $z_{a}=0$ and $z_{b}=1$, we have $\mathcal{N}(b)=a$, which implies that the configuration $2$ will appear in the graph $\mathfrak{G}$. The same argument is valid for the case of $Z_{a}=1$ and $z_{b}=0$. When $z_{a}=z_{b}=1$, Eq. (\ref{eqsu}) requires that the vertices $a$ and $b$ have the same neighbours besides themselves. Namely, if the two vertices are not neighbours, it leads to the configuration $1$ in the graph $\mathfrak{G}$; otherwise, it leads to the configuration $2$ in the graph $\mathfrak{G}$. This completes our proof of the above observation.   

Second, we find that if there are two elements whose supports' dimension are $2$ in the stabilizer group of the graph state $|\Psi_{\mathfrak{G}}\rangle$, then either the two supports are disjoint, or the intersection of the support is a set with one vertex, and the operators of the vertex for the two elements are the same.

This proposition can be shown as follows. Let us denote the two elements as $O_{1}^{(a)} O_{1}^{(b)}$ and $O_{2}^{(c)}O_{2}^{(d)}$. When their supports are the same, \textit{e.g.}, $a=c$ and $b=d$, the operators of the same vertex are not the same, otherwise they are the same element from the community of the two elements. If the operators of the same vertex are not the same, then the state for these two vertices is the maximally entangled state, which implies that the two vertices are not connected with the other vertices in the graph. This leads to a contradiction to our assumption of a connected graph we consider. When the intersection of the supports of the two elements contains only one vertex, then the operators of the vertex for these two elements must be the same, otherwise the two elements are not commutative. This completes our proof of the second proposition.

Third, we concludes that the stabilizer dimension of the graph state $|\Psi_{\mathfrak{G}}\rangle$ is equal to the cardinality of $\mathfrak{g}_{2}$ of the state $|\Psi_{\mathfrak{G}}\rangle$, which is equal to the degree of irreducible two-qubit correlation for this state.  

This conclusion is obtained based on the discussions given above. The stabilizer dimension of the graph state $|\Psi_{\mathfrak{G}}\rangle$ is equal to the number of the independent generators of Lie algebra associated with the three configurations appearing in the graph $\mathfrak{G}$. Based on the first proposition, we know that every element in the $\mathcal{G}_{2}$ also leads to the same configurations. This implies that a one-to-one map can be built between the elements in $\mathcal{G}_{2}$ and the elements in the Lie algebra by the rules:
\begin{equation}
O^{(a)}O^{(b)}\rightarrow O^{(a)}-O^{(b)}.
\end{equation}
According to the proposition $2$, we find the number of independent elements in $G_2$ is equal to the number of independent elements in the Lie algebra by the above map. In Ref. \cite{Zho08}, we proved that the degree of irreducible two-qubit correlation is equal to the cardinality of $\mathfrak{g}_{2}$. Therefore, we complete the proof of our conclusion.         

\section{Discussion and summary}

According to the definition of the graph state, we know that all the elements in the stabilizer group, which are local unitary transformations, stabilize the state. However, our findings show that only the elements in the stabilizer group whose supports' dimension is $2$ contribute to the stabilizer dimension of the state. This implies that the elements in the stabilizer group are not equivalent for local unitary transformations. However, all the generators play the same role in generating the group. One possible solution to this puzzle is to study the more general unitary transformations that stabilize the state. It might be a future topic along this direction.

In summary, we obtain the stabilizer dimension for arbitrary connected graph states. We find that the stabilizer dimension for a graph state is associated with three specific configurations in the graph. We further show that the stabilizer dimension for an $n$-qubit ($n\ge 3$) graph state is equal to the cardinality of $\frak{g}_{2}$, which is the degree of irreducible two-qubit correlation as shown in Ref. \cite{Zho08}. We hope that our results would shed new light on the characterization of multiparty entanglement in a multiparty quantum state.

This work is supported by NSF of China under Grant No. 10775176, and NKBRSF of China under Grant Nos. 2006CB921206 and 2006AA06Z104.

\end{document}